\documentstyle[12pt]{article}
\begin{document}
\baselineskip .3in
\begin{titlepage}
\begin{center}{\large{\bf Properties of Proton in diquark Model}}
\vskip .2in A.Bhattacharya$^{\dag}$ ,
S.N.Banerjee,B.Chakrabarti,S.Banerjee
\end{center}
\vskip .1in
\begin{center}
Department of Physics, Jadavpur University \\
Calcutta 700032, India.\\
\end{center}
\vskip .3in

\vskip .3in Abstract: The properties of the proton has been
investigated considering a proton as diquark-quark system. A model
for diquark has been suggested in an analogy with the
quasi-particle in a crystal lattice. The mass of the diquark
obtained in the present model has been found to be in good
agreement with other theoretical predictions. The binding energy
of proton, compressibility and excitation energy for the Roper
Resonance have also been estimated in the present model and are
found to be in agreement with the existing theoretical and
experimental findings.

Pacs: 14.20-c; 14.65-q; 14.20Dh; 12.39Pn


\vskip 1in

$^{\dag}$e-mail: pampa@phys.jdvu.ac.in
\end{titlepage}
\newpage
Diquarks play an important role in the structure of hadrons. It is
now growing idea that the deeply bound diquarks are the building
blocks in the formation of the meson and baryon states and the
exotics. Some experimental facts like hadron jet formation etc
demands the prediction of diquark and the role of diquark in
baryon spectroscopy, deep-inelastic function and dynamics has been
discussed in details by Ida et al.and Anselmino et al [1].
Particularly after the discovery of the pentaquark baryon
$\theta^{+}$ [2], the search on diquark attract much attention. A
number works have been done towards the understanding of the
structure of diquark. The possibility of forming quark-quark and
quark-antiquark system by Instanton Induced Interaction(III) have
been developed by Shuryak [3] and Schafer et.al [4]. Betman et.al.
have investigated formation of bound state of quark-quark or
quark-antiquark systems due to instanton induced interaction and
predicted such bound state in the hadron as a bubble of the size
of the instanton radius. In QCD both gluon exchange interaction or
instanton induced interaction favour spin singlet colour-
antisymmetric diquark combination. The discovery of the exotic
baryon $\theta^{+}$ which has the mass much less than the
constituent quark, the diquark picture in the framework of III
model reproduces much satisfactory result. A number of works have
been done on the diquark picture of the pentaquark [5]. If diquark
truely exists and proved to be a fundamental building block of
hadrons, it must also reproduce the nucleon or other baryon
properties as well.
     In the present work we have suggested an alternative picture
     of diquark in which two quarks bound together forming a quasi
     particle in analogy with the quasi particle formed in the
     crystal lattice [6]. We have estimated the diquark mass in
     the framework of the quasi-particle in the lattice and the
     results are obtained is in good agreement with the other
     theoretical predictions. We have estimated the proton binding
     energy, compressibility and excitation energy of the Roper
     resonance,the first excited state of the nucleon in this quasi-particle
     model and obtained favourable results.

The electron in crystal is in a situation exactly the same as the
elementary particle in vacuum [7]. The force will be acting on
particle which are already under crystal field. An crystal
electron is subjected to two force namely the effect of crystal
field ($\nabla V$) and an external force (F) which accelerate the
electron. So the electron in a crystal behaves like a quasi
particle whose effective mass $m^{*}$ reflects the inertia of
electrons which are already in a crystal field such that:
\begin {equation}
m^{*}\frac{dV}{dt}= F_{h}
\end{equation}

and the bare electrons ( with normal mass) are affected by the
lattice force -$\nabla V$,  which corresponds the periodic crystal
potential V as well as the external force F. so that:
\begin{equation}
m\frac{dV}{dt}= F - \frac{dV}{dx}
\end{equation}
 Hence the ratio of the normal mass (m) to the effective mass ($m^*$) can
expressed as:
 \begin{equation}
 m/m^{*}= 1 - \frac{1}{F}[\frac{\delta \overline{V}}{\delta x}]
 \end{equation}
We propose a similar type of picture for diquark as quasi particle
inside a nucleon.
 To get diquark effective mass inside the nucleon we assume that the
 diquark is an independent body which is under the influence of one
 gluon exchange type of field due to the meson cloud represented by potential
 $\overline{V}$
 =-(4/3)$\alpha_{s}$/r in analogy with the crystal field on a crystal
 electron.
 and an average of lattice force F = -ar as  an oscillatory
 external force so that the the ratio of the constituent mass and the
 the effective mass of the diquark inside a hadron may be
 represented as:
 \begin{equation}
 \frac{m}{m^{*}}= 1+ \frac{\alpha_{s}}{ar^{3}}
 \end{equation}
 Here m represents the normal constituent mass of the diquark and
 m$^{*}$is the effective mass of the diquark,$\overline{V}$ being the average value of the one gluon
 exchange potential.
 To calculate the effective mass of the diquark we need the
 parameter r for the diquark. In the context of discussing the diquarks in
 instanton induced interaction model Betman et.al. [5] have
 considered the instanton spacial distribution function F(r) as
 F(r) =$\delta$(r-$\rho_{c}$ where $\rho_{c}$ is the characteristic
 instanton size which also represents radius of the bound state of diquark
 in the form of bubble of the size of the instanton radius. We have assumed that the'r'
 parameter for the diquark is of the order of $\rho_{c}$ and taken to be = 0.4 fm as
 in [5]. With $\alpha_{s}$ = .59 [8] for light hadrons and r = 0.4
fm, m =.72GeV ($m_{u}$= .36GeV) and  a = .06 [9], we get the
effective mass of the diquark as 272 MeV in the quasi particle
model of the diquark. Jaffe et al. [10] estimated the diquark mass
420 MeV in the context of III interaction whereas Kl Model [11]
estimated the mass to 209 MeV. Some other calculations [5,12]
estimated the mass in the range between 600 to 800 MeV.

To estimate the binding energy of proton we consider ud-u picture
of proton where (ud$_{0}$) constitutes the diquark of mass 272
MeV. We then estimated the reduced mass of the diquark-quark
(M$_{R}$) system and consider that a particle of mass M$_{R}$ is
moving under a potential V(r) = ar$^{2}$ with the centre of mass
fixed to the centre of the baryonic sphere. The reduced mass is
estimated  to be as 0.155 GeV. It may be mentioned here that
Krolikwoski [13] has made one body equilateral triangle
approximation to the three(Dirac) particle system for estimation
of the Hamiltonian. Here the expression for the Hamiltonian
corresponding to the particle of mass M$_{R}$ moving in a
background potential V(r)= ar$^{2}$ due to the sea contribution
runs as:
 \begin{equation}
 H = -\frac{\hbar^{2}}{2M_{R }}+ ar^{2}
  \end{equation}
  The expectation value of H corresponding to the wave function
  $\psi(r)$ is;
  \begin{equation}
<H> = \int \psi^{*} H \psi d^{3}r
\end{equation}
 The lowest upper bound of the ground state energy for a wave function $\psi$ may be
recast as:
\begin{equation}
 E_{0}\leq  \int \psi^{*} H \psi d^{3}r
 \end{equation}
 In the context of the Statistical model [14] the square of the wave function of a
 proton for harmonic oscillator type of background potential is
 obtained as:
 \begin{equation}
 \psi(r)= A^{\frac{1}{2}}(r_{0}^{2}-r^{2})^{\frac{3}{4}} e^{i\alpha}\theta(r_{0}-r)
 \end{equation}
 Where $r_{0}$ corresponds to the radius parameter of a proton. $\theta$ is usual step
 function.$\alpha$ is a constant phase factor, A =
 $\frac{8r_{0}^{-6}}{\pi^{2}}$[13]
 From the above expression we have the |$\nabla\psi$| and
 corresponding $<T>$ has been estimated with the above baryonic
 wave function as:
 \begin{equation}
 <T> = 43.49 r_{0}^{-2}
 \end{equation}
 The expectation value of the potential energy has been estimated
 as:
 \begin{equation}
 <V> = 0.375a r_{0}^{2}
 \end{equation}
 Hence the total energy runs as:
 \begin{equation}
 <E> = 43.49r_{0}^{-2}+ 0.375a r_{0}^{2}
 \end{equation}
 $r_{0}$ is typical radius parameter of the proton and can be
 estimated using the experimental data corresponding to form
 factor. The form factor of proton has been obtained from the
 expression $F(q^{2})$ = $\int e^{iq.r}|\psi(r)|^{2}d^{3}r$. With
 the above $|\psi(r)|^{2}$ as input we obtain for
 $q^{2}\longrightarrow 0$;
 \begin{equation}
 F(q^{2})= 1- 0.057r_{0}^{2}q^{2}
 \end{equation}
 The relation between the proton charge radius and form factor for
 a proton is:
 \begin{equation}
 F(q^{2}) = 1- \frac{1}{6}<r_{ch}^{2}>q^{2}
 \end{equation}
 Comparing above expressions for $F(q^{2})$ we obtain;
 $<r_{ch}^{2}>$ = 0.6$r_{0}$.
  The proton charge radius is obtained as [15]
 $<r_{ch}^{2}>^{\frac{1}{2}}$ = 0.88fm which yields $r_{0}$=
 1.46fm = 7.33GeV. With the input of this radius we obtain $<E>$ =
 1.927 GeV. Lim [16]  has solved the three body (trinucleon) problem
 with the spin dependent internuclear harmonic oscillator
 potential and obtained the binding energy of the triton as:
 \begin{equation}
 E =(\frac{2\hbar^{2}}{m})^{1/2}\{(3\omega V_{1} +
\frac{3}{2}b_{1}V_{1})^{1/2})+(3\omega V_{1} -
  \frac{3}{2}b_{1}V_{1})^{1/2}\} -3\omega V_{0}
  \end{equation}
  where $V_{0}$ $V_{1}$, $b_{1}$, $\omega$ are appropriate
  constants. Considering the three body system as the three quark
  system with mass of m as $m_{q}$ (360 MeV),quark mass, the binding energy
  for ground state of proton is obtained as $<E>$ = 1.997 GeV
  (with $\frac{1}{2}Kr^{2}$= $V_{2}r^{2}$ = $V_{1} \omega r^{2}$ =
  a$r^{2}$.
  So we find that the our estimate of the binding energy for
  proton in the context of the quasi-particle model agree closely
  with the Lim's exact theoretical calculation of the ground state
  energy.

    It has been pointed by Morsch et al. [17]
   that the information on the compressibility of a system can be obtained from the
   dynamical properties of the size degree of freedom in radial
   mode.The compressibility of a nucleon is given by the expression;
   \begin{equation}
   K = r_{0}^{2} \frac{1}{3}(d^{2}E/dr^{2})
   \end{equation}

     To get an estimate of the compressibility of nucleon we use
    the radius parameter of nucleon = 7.33GeV as above
     and get the value as $K_{N}$ = 1.61GeV
     The Roper excitation energy is given by:
    \begin{equation}
    \Delta E = \frac{K_{N}}{m_{q}r_{0}^2}
    \end{equation}
     We estimated the Roper excitation energy as 288 MeV in the
     present work.
      However it may be mentioned that the Roper resonance form factor
     is smaller than proton form factor which indicates the fact that the
    Roper being a more diffuse system than proton[18]. Morsch et al.[17]  have investigated
    $P_{11}$(1440) in the alpha-photon
    scattering. With the mean quare radius equals to 0.62$fm^{2}$, they
    have extracted the $K_{N}$ = 1.4 $\pm$ 0.3GeV. MIT [19]  Bag model
    predicts the value in between 900 to 1200 Mev whereas the
    constituent quark model yields the value as $\simeq$ 3Gev.
    Mathieu et el [20] have estimated the value as 636 MeV in the
    context of the flux tube model. Hoodbhoy et.al [21] have
    investigated the pion mediated interaction in a chiral bag
    model considering nucleon size and compressibility as
    a parameter. They have taken $K_{N}$ as 2GeV and 3GeV
    with radius parameter as .6fm to.8fm. Meissner et.al.[22] estimated
    the  compressibility as 4GeV.

    We have estimated the excitation energy of Roper resonance as 288 MeV with $K_{N}$ = 1.61 GeV
    as estimated by us. Meissner et al.[22] have estimated the excitation energy to be 390
    Mev. However they have mentioned that all the relativistic
    approaches estimate the roper resonance excitation energy
    ranging from 200MeV to 500MeV whereas the experiment predicts
    the value as 500Mev.

               In the present work we have suggested  a quasi particle picture of diquark
               which resembles a quasi particle in a crystal
               lattice. It may be pertinent to mentioned here that
               in the context of investigating the super conducting  properties of a
               hadrons [23], it has been suggested that vacuum is
               supposed to contain a sea of virtual
               q$\overline{q}$ pairs condensates somewhat similar to the
               situation of Cooper pairs in a superconductor. The
               particle picture description of hadrons may have
               some relevance to the recently developed idea that
               diquarks are the building blocks of hadrons and
               exotics. The binding energy we obtain does not mean
               the binding energy in usual sense as quarks are
               not free but binding energy should be more than the
               mass of the proton. We have observed that if a
               trinucleon system is replaced by three quarks to
               represent a proton
               the binding energy obtained from the exact solution
               of Lim [16]
               is almost equals to our calculation obtained from  the quasi
               particle picture of diquark-quark of proton. The compressibility of
               the nucleon estimated in the present work lies in
               the range of recent extracted value [17]. The Roper
               excitation energy is obtained in the range of other
               theoretical estimates. It should be mentioned that
               most uncertainty lies in the estimation of the
               radius parameter which is not very well known [17].
               In the context of discussing the pentaquark baryons
                Oka [6]  has pointed out that situation is
               not very clear with the diquark picture but
               diquarks if exists should be reexamined with the
               other ground state baryons. The model we have
               suggested for diquaks reproduces the
               properties of proton in existing theoretical and experimental limits
               and may not be far from reality. However further
               investigations would be made with the
                exotics particularly with the pentaquark
               system in our future works.

\newpage

\noindent {\bf REFERENCES}

\noindent [1] M.Eda et al.,  Prog. Theor.Phys.{\bf 36}, (1966)846;
                  M.Anselmino et al. Rev Mod. Phys
                  {\bf65},(1993)1199

\noindent [2] T.Nakano et al., Phys. Rev. Lett. {\bf 91},(2000)
262001

\noindent [3]E.V. Shuryak., Nucl. Phys. {\bf B50},(1982)93

\noindent [4] T. Schafer et al., Rev. Mod. Phys.{\bf 70},(1998)323

\noindent [5] R.G. Betman et al.,Sov.J.Nucl.Phys{\bf 41},
(1985)295

\noindent [6] M.Oka., Prog. Theor. Phys.{\bf 112}, (2004)1; R. L.
Jaffe., Phy. Rep.{\bf 409},(2005)1

\noindent [7] A. Haug., Theoretical Solid State Physics,Pergamon
Press; (1972)100

\noindent [8]. W. Lucha et al., Phys.Rep.{\bf 200},(1991)127

\noindent [9] M. Hirano et al., Prog. Theor. Phys.{\bf 35},
(1971)645

\noindent [10] R.L. Jaffe and F. Wilczek.; Phys. Rev.Lett.{\bf
91},(2003)55

\noindent [11] M. Karliner and H. J. Lipkin., Phys. Lett.{\bf
B575},(2003)249

\noindent [12]Y. Kanada-En'yo et al.,Hep-ph/0404144, P. Maris.,
Few Body. Syst.(2002)32; Nucl-th/0204020

\noindent [13] W. Krolikwosski., Phy. Rev.{\bf D31},(1985)2659

\noindent [14]  A. Bhattacharya et al., Prog. Theor. Phy. {\bf
77}, (1987)16

\noindent [15] Particle Data Group., Phys. Lett.{\bf B592},
(2004)854

\noindent [16] T.K. Lim., Am. J. Phys. {\bf 39},(1971)932

\noindent [17] H.P. Morsch et al., Phys. Rev. Lett {\bf
69},(1992)1336; Z. Phys.{\bf A350},(1994)61

\noindent [18]G.L. Strobel et al., Int. J. Theor. Phys {\bf 38},
(1999)737

\noindent [19] R. K. Bhadhuri et al., Phys. Lett. {\bf B136},
(1984)289

\noindent [20] P. Mathieu et al., Can J. Phys {\bf 64}, (1986)1389

\noindent [21] P. Hoodhoy et al., Phy. Rev. {\bf C28},(1983)1455

\noindent [22] T. Meissner et al., Phys. Rev. {\bf D39},(1989)39

\noindent [23] D.E. Ebrt et al., Fortsch. der Physik.
{\bf},(1989)487; S.N. Banerjee et.al., Int. J Mod. Phys.{\bf
17},(2002)4939

\end{document}